\documentclass[a4paper,fleqn,final]{cas-dc}

\usepackage[T1]{fontenc}
\usepackage[utf8]{inputenc}
\usepackage[main=english]{babel}
\usepackage{textcomp}
\usepackage{hyperref}
\usepackage{amsmath,amssymb}
\usepackage{graphicx}
\usepackage{booktabs}
\usepackage{tabularx}
\usepackage{array}
\usepackage{calc}

\usepackage{threeparttable}
\usepackage{longtable}
\usepackage{adjustbox}
\usepackage[numbers,sort&compress]{natbib}
\usepackage{caption}
\usepackage{capt-of}
\usepackage{etoolbox}
\usepackage{placeins}
\usepackage{stfloats}
\usepackage{cuted}
\usepackage{balance}
\usepackage{needspace}
\usepackage{xurl}
\usepackage{microtype}
\usepackage{lastpage}

\extrafloats{100}
\newcommand{\reviewfig}[1]{\includegraphics[width=\textwidth]{#1}}
\newcommand{\tablenote}[1]{%
  \par\vspace{0.25\baselineskip}%
  \noindent\begin{minipage}{\linewidth}%
  \footnotesize #1\par
  \end{minipage}\par
}

\makeatletter
\g@addto@macro{\UrlBreaks}{\do\_\do\-\do\/\do\.\do\0\do\1\do\2\do\3\do\4\do\5\do\6\do\7\do\8\do\9}
\makeatother

\renewcommand{\ttfamily}{\rmfamily}
\AtBeginDocument{%
  \urlstyle{same}%
}

\usepackage{xcolor}
\definecolor{linknavy}{RGB}{0,70,127}
\hypersetup{
  colorlinks=true,
  linkcolor=black,
  citecolor=linknavy,
  urlcolor=linknavy,
  runcolor=linknavy
}

\AtBeginDocument{%
  \raggedbottom
  \setlength{\abovedisplayskip}{5pt plus 2pt minus 1pt}%
  \setlength{\belowdisplayskip}{5pt plus 2pt minus 1pt}%
  \setlength{\abovedisplayshortskip}{2pt plus 1pt}%
  \setlength{\belowdisplayshortskip}{4pt plus 1pt minus 1pt}%
}

\AtBeginEnvironment{table}{\let\sffamily\rmfamily}
\AtBeginEnvironment{figure}{\let\sffamily\rmfamily}
\AtBeginEnvironment{table*}{\let\sffamily\rmfamily}
\AtBeginEnvironment{figure*}{\let\sffamily\rmfamily}

\ExplSyntaxOn
\RenewDocumentCommand \firstname {}
  { \textcolor{black}{\seq_use:Nn \l_stm_au_seq { ~ }} }

\RenewDocumentCommand \emailauthor { m m }
   {
     \int_gincr:N \g_ead_int
     \seq_gput_right:Nn \g_stm_ead_seq
       {
         { \href{mailto:#1}{\rmfamily #1} }
         \parsename { #2 }
         \space(\eadauthor)
       }
     }

\RenewDocumentCommand \printemails { }
{
  \group_begin:
  \int_compare:nNnTF { \int_use:N \g_ead_int } > { 0 }
  {
    \tex_let:D \thefootnote \relax \footnotetext
    {
      \raggedright
      \raisebox{-0.5pt}{\includegraphics[height=7.5pt]{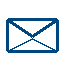}}\c_space_token
      \seq_use:Nn \g_stm_ead_seq { ;~ }
    }
  }
  {  }
  \group_end:
}

\cs_set:Npn \__first_footerline:
{
  \group_begin:
  \small
  \normalfont
  \ifnum\theblind>0\relax
  \else
  \__short_authors: :~
  \fi
  \itshape Preprint~submitted~to~Frontiers
  \group_end:
}

\RenewDocumentCommand \printorcid { } { }

\cs_set:Npn \__first_head:
{
  \parbox[t]{\textwidth}
  {
    \rule{\textwidth}{0pt}
  }
}

\cs_set:Npn \__cas_head:
{
  \parbox{\textwidth}
  {
    \rule{\textwidth}{0pt}
  }
}

\cs_set:Npn \__cas_foot:
{
  \parbox[t]{\textwidth}
  {
   \rule{\textwidth}{.2pt}\\
   \small
   \normalfont
   \__first_footerline:
   \hfill Page~\thepage {}~of~ \lastpage
  }
}
\ExplSyntaxOff

\makeatletter
\ps@cas
\makeatother

\begin{document}

% IEEE BST control: show full author names (no ------ dashes)
\makeatletter
\def\bstctlcite{\@ifnextchar[{\@bstctlcite}{\@bstctlcite[@auxout]}}
\def\@bstctlcite[#1]#2{\@bsphack
  \@for\@citeb:=#2\do{%
    \edef\@citeb{\expandafter\@firstofone\@citeb}%
    \if@filesw\immediate\write\csname #1\endcsname{\string\citation{\@citeb}}\fi}%
  \@esphack}
\makeatother
\bstctlcite{IEEEbsTcontrol}

\shortauthors{Farajpoor and Narimani}
\shorttitle{GeoAI for climate resilience}

\title[mode=title]{Geospatial foundation models and deep learning for climate resilience: a review of transfer, uncertainty and decision readiness}

\author[1]{Parastoo Farajpoor}
\author[1]{Mohammadreza Narimani}
\cormark[1]
\ead{mnarimani@ucdavis.edu}

\affiliation[1]{organization={Department of Biological and Agricultural Engineering, University of California, Davis},city={Davis},state={CA},postcode={95616},country={USA}}

\cortext[1]{Corresponding author}

\begin{abstract}
Geospatial foundation models can reuse information learned from large Earth-observation archives, but their value for climate resilience depends on whether that information transfers to unfamiliar events, regions and observing conditions. This review connects advances in deep learning with the evidence needed for hazard monitoring and early warning. A descriptive audit of 137 full-text hazard-application studies is interpreted alongside a separate corpus of 37 model and benchmark records. Ninety-eight application studies reported an out-of-distribution experiment, whereas headline evaluations included held-out events in 31 studies, regions in 25 and sensors in only two. None of the ten studies coded for geospatial foundation-model fine-tuning used sensor holdout as a headline test. Uncertainty-related methods were recorded in 16 studies, including two with explicitly calibrated probabilistic output. These overlapping counts describe evaluation practice within the coded corpus, not a ranking of model families. We distinguish Earth-observation representation learning from atmospheric forecasting and examine how labels, observability, forecast lead time and decision requirements constrain their contribution across flood, wildfire, drought, heat, cryosphere and cascading-hazard applications. We then develop a decision-backward evaluation framework: define the action and information available when it must be taken, reproduce the relevant shift, retain a matched task-specific baseline, and report misses, false alarms, reliability and end-to-end latency. The framework turns transferability from a broad model claim into a testable requirement for climate adaptation and risk management.
\end{abstract}


\maketitle

\section{Introduction}
\label{sec:1}
A flood map delivered after inundation and a warning issued before peak discharge can both be accurate and still serve very different purposes. The first supports inspection and recovery; the second must arrive early enough to change an action. Their usefulness depends not only on a model's predictions, but also on what was observable, when information became available and which mistakes matter to the user. This distinction is central to applying geospatial artificial intelligence (GeoAI) to climate resilience. A model that recognizes familiar landscapes is not necessarily a model that remains reliable during an unfamiliar extreme.

Deep learning has expanded the use of satellite imagery, environmental time series and physical-model output for Earth-system analysis. Geospatial foundation models extend this development by pretraining representations on large, heterogeneous archives and adapting them to downstream tasks. In parallel, machine-learning weather models predict the evolution of atmospheric fields. These developments create opportunities for climate monitoring and adaptation, but their evaluation must connect computational skill with an environmental process and a practical decision \citep{reichstein19,camps,reichstein25}.

Transfer is particularly important for hazards because the next event is not a random new image from an unchanged distribution. It may occur in a different basin, biome or built environment; coincide with persistent cloud; or require a sensor and processing chain that were absent during development. Reference labels also vary: surface water, newly inundated land, active fire, burned area and structural damage are different targets. Generalization therefore has several dimensions, and a high score under one split cannot establish performance under all of them. Spatially and temporally structured validation provides the methodological foundation for making these distinctions \citep{roberts,meyer18}.

Existing reviews have explained AI applications to extremes, the progression of satellite hazard-monitoring methods and the opportunities for integrated early-warning systems. Benchmark studies have also shown why apparently comparable foundation-model scores can depend on evaluation settings. Our contribution is complementary: we examine the evaluation designs used in hazard applications and connect them to the decisions that a deployment must support. Table~\ref{tab:tab1} positions this contribution without treating architectural novelty, broader pretraining or a particular model family as evidence of operational readiness \citep{bentivoglio,illarionova,corley}.

\begin{table*}[!t]
\centering
\small
\caption{How this review complements related syntheses}
\label{tab:tab1}
\begin{tabular}{p{0.22\textwidth}p{0.35\textwidth}p{0.35\textwidth}}
\toprule
\textbf{Synthesis} & \textbf{Main contribution} & \textbf{Connection to the present review} \\
\midrule
Flood mapping review (Bentivoglio et al., 2022) & Deep-learning applications, data and open problems in flood mapping. & Connects the observation and mapping task to its validation requirements. \\
AI for extremes (Camps-Valls et al., 2025) & Detection, forecasting, understanding and impacts across extreme events. & Provides broader Earth-system context for the application-level audit. \\
Satellite hazard monitoring (Illarionova et al., 2026) & Development of satellite-based AI methods and their environmental applications. & Places newer representations alongside established hazard-monitoring methods. \\
Integrated early warning (Reichstein et al., 2025) & Links meteorological, geospatial and decision-system capabilities. & Motivates issue-time verification of coupled weather-to-impact systems. \\
Foundation-model benchmark audit (Corley et al., 2026) & Examines consistency and reproducibility of model evaluation reports. & Complements the present focus on hazard-specific headline transfer tests. \\
\bottomrule
\end{tabular}
\tablenote{The contributions are complementary. The present review adds descriptive counts for a fixed 137-study application corpus and a decision-backward testing framework; it does not infer that the cited syntheses exclude all of these themes.}
\end{table*}

The review addresses three questions. Which shifts are actually tested in the coded hazard literature? Which constraints can representation learning plausibly relieve, and which require better observations, process information or decision design? What controlled experiments would establish a useful contribution from pretraining? We answer these questions through a fixed descriptive audit, a synthesis of relevant model and application studies, and a decision-backward protocol. The emphasis is climate adaptation and risk management: improvements must be judged against the information and resources available to the people expected to act on them.

\section{Review approach and interpretation}
\label{sec:2}
\subsection{Evidence sources and corpus boundaries}
\label{sec:21}
The application corpus was assembled through 35 OpenAlex title-and-abstract queries covering publications from 2015 to 2026 (search date 1 November 2026), with up to 200 relevance-ranked records retrieved per query under the anonymous daily API budget. The query set spanned foundation-model and self-supervised remote-sensing terms together with flood, wildfire, drought, heat, cryosphere, compound-hazard, damage-mapping, forecasting and evaluation families. Exact query strings, reported hit counts and retrieved counts are retained in the project search log and are available from the corresponding author upon reasonable request. Retrieval yielded 6,803 raw records, 6,092 unique DOI/title keys after deduplication, and 5,867 records after near-duplicate title-twin removal. Automated title prescreening and human title--abstract screening retained 141 primary candidates spanning flood, wildfire, drought, heat, cryosphere, damage assessment and compound or cascading hazards. Evaluation-design coding used the 137 records with usable full text. This is a structured critical review with a descriptive evidence audit, rather than a pooled meta-analysis of predictive accuracy or a PRISMA systematic review.

A separate collection of 37 model and benchmark records provides architectural context: 18 Earth-observation foundation models, nine weather or climate models, five evaluation benchmarks and five benchmark datasets. These records are not added to the denominator of 137 application studies. Literature consulted to update the narrative and verify references through 22 November 2026 informs the synthesis but does not retrospectively alter the audit counts. This separation keeps the descriptive results stable while allowing discussion of recent developments.

\subsection{Coding and units of analysis}
\label{sec:22}
Application studies were coded from usable full text under a fixed coding protocol. Each record received codes for hazard, task, modality, model family, pretraining, study geography (locations, never affiliations), headline validation design, additional out-of-distribution tests, baselines, uncertainty-related methods, explainability, and code or weight availability. The unit of analysis is the study record, not the image chip, model run or country mention. A record can carry several hazard, model and validation tags. Coders worked only from the study text: when a validation description was ambiguous, the more conservative code was assigned; abstract-only evidence was not used to infer evaluation designs. Study-level coding tables for the 137 usable application records (from 138 verified) and the parallel 37-record foundation-model and benchmark extraction are available from the corresponding author upon reasonable request.

The headline validation code describes the test supporting a study's principal reported evaluation. An additional out-of-distribution experiment can occur elsewhere in the same paper. Consequently, the indicator 'any out-of-distribution test' and the counts of specific headline holdouts answer different questions. Code availability denotes an explicit availability statement or link in the coded paper, not a continuing guarantee that a repository remains accessible. An uncertainty-related code records a reported method; it does not by itself establish calibration, coverage or suitability for a decision.

Table~\ref{tab:tab2} defines the main contrasts used in the synthesis. For sensor change, the distinction between an unseen sensor, a missing modality and adaptation to a new input configuration is especially important. These settings may all be operationally relevant, but they do not establish the same form of transfer. Similarly, downstream test separation does not establish that a location or acquisition period was absent from a foundation model's pretraining archive.

\begin{table*}[!t]
\centering
\small
\caption{Evaluation settings and the claims they can support}
\label{tab:tab2}
\begin{tabular}{p{0.22\textwidth}p{0.35\textwidth}p{0.35\textwidth}}
\toprule
\textbf{Setting} & \textbf{Required separation or control} & \textbf{Interpretation} \\
\midrule
Random split & Samples are divided within the sampled data distribution. & Addresses within-distribution performance; geographic or event independence is not implied. \\
Spatial block & Contiguous spatial units are withheld. & Tests spatial separation at the selected block scale; distance and residual dependence matter. \\
Held-out time & Target times are excluded from fitting, tuning and calibration. & Supports a temporal claim when all predictors were available at the stated issue time. \\
Held-out event & All observations from the test event are excluded from downstream development. & Tests a new event; multiple independently held-out events establish breadth. \\
Held-out region & A defined geographic domain is withheld from downstream development. & Tests geographic transfer at that scale; environmental similarity still matters. \\
Sensor transfer & A new sensor or input configuration is tested with an explicit adaptation budget. & Distinguish unseen-sensor evaluation, missing-modality tests and target-sensor adaptation. \\
Pretraining independence & Document whether test locations, times or sensors occur in the pretraining archive. & Downstream holdout alone does not establish independence from pretraining. \\
\bottomrule
\end{tabular}
\tablenote{These settings can overlap. Descriptions concern the claim supported by a design, not a universal ranking of test difficulty. General methodological basis: \citep{roberts,meyer18}.}
\end{table*}

\subsection{Descriptive synthesis}
\label{sec:23}
Unless a different denominator is stated, percentages are calculated as 100 times the number of coded studies divided by 137. Model-group comparisons use ten foundation-model fine-tuning studies and 127 other studies. Hazard-specific figures use the corresponding hazard subset. Cross-tabulations count records carrying both tags; sums across hazards or model families are therefore not counts of distinct papers. No significance test is used to claim that one model group has a better evaluation culture than another, and no cross-paper accuracy average is calculated.

Study-country coverage is based on reported locations, not author affiliations. Of the 137 application records, 110 have eligible country codes; the coding contains 86 ISO3 territories. A multi-country study contributes once to each coded territory, and training and evaluation areas may both contribute. The resulting map characterizes the geography of the evidence rather than hazard incidence or independently validated deployment coverage. Aggregate values and coding definitions accompanying the figures are available from the corresponding author upon reasonable request.

\section{From observations to decisions}
\label{sec:3}
\subsection{Separate the product from the action}
\label{sec:31}
\begin{figure*}[!t]
\centering
\reviewfig{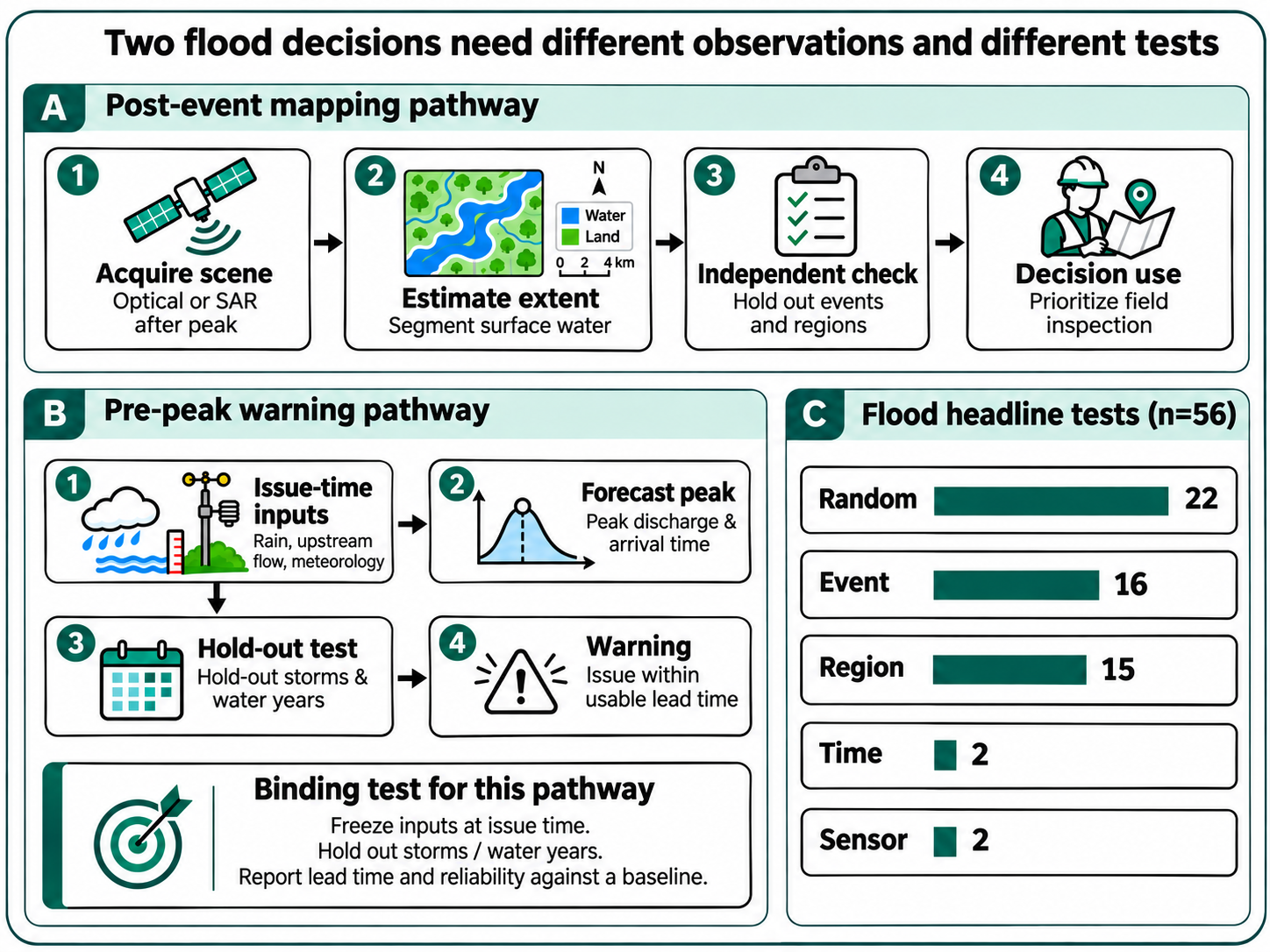}
\caption{Two flood decisions need different observations and different tests. (A) Post-event mapping pathway. (B) Pre-peak warning pathway. (C) Headline validation designs among flood studies in the usable audit.}
\label{fig:fig1}
\end{figure*}

Figure~\ref{fig:fig1} contrasts two flood pathways. In post-event mapping, an optical or synthetic aperture radar (SAR) acquisition is converted into a surface-water or inundation product and checked against independent events or regions. The result can prioritize field inspection. In pre-peak warning, the inputs must be those available at the forecast issue time: rainfall, upstream flow, meteorology and other admissible antecedent information. A warning is useful only if it meets the required lead time and has acceptable reliability. A retrospective map, even a highly accurate one, cannot by itself demonstrate this forecasting capability.

The task definition determines what counts as a valid label. Water observed at satellite overpass is not necessarily maximum flood extent, and permanent water is not newly flooded land. Acquisition timing, cloud, canopy, topography and mixed pixels can change the relationship between a label and the event of interest. Reviews of flood mapping and cloud-aware optical mapping systems show why sensor characteristics and reference construction must remain part of the evaluation, rather than being treated as preprocessing details \citep{bentivoglio,portales}.

\subsection{Match the observation to the process}
\label{sec:32}
The coded studies most frequently used multispectral optical imagery (70), digital elevation models or static covariates (54), SAR (46), reanalysis or meteorological information (34), in situ observations (29), coarse-resolution optical data (29), very-high-resolution optical imagery (28), and thermal infrared data (21). These are overlapping modality tags. Their diversity reflects different measurement problems: an active-fire signal, a canopy-temperature estimate and a snow-water-equivalent estimate are not alternative labels for the same phenomenon.

For an operational comparison, all models should receive a clearly defined information set. A model using more bands, later observations or additional local labels has an informational advantage that should not be attributed entirely to its architecture. In H2O-Net's PlanetScope evaluation, the RGB model achieved 60.92\% mean intersection over union (mIoU), compared with 42.63\% for an RGB U-Net. The H2O-Net result increased to 74.52\% with near-infrared added. The matched RGB comparison tests methods; the additional-band comparison changes the information available to the method \citep{akiva}. This distinction is simple, but essential when interpreting claims of multimodal or foundation-model superiority.

\section{Foundation models: capabilities and evidence boundaries}
\label{sec:4}
\subsection{Earth-observation representation learning}
\label{sec:41}
\begin{figure*}[!t]
\centering
\reviewfig{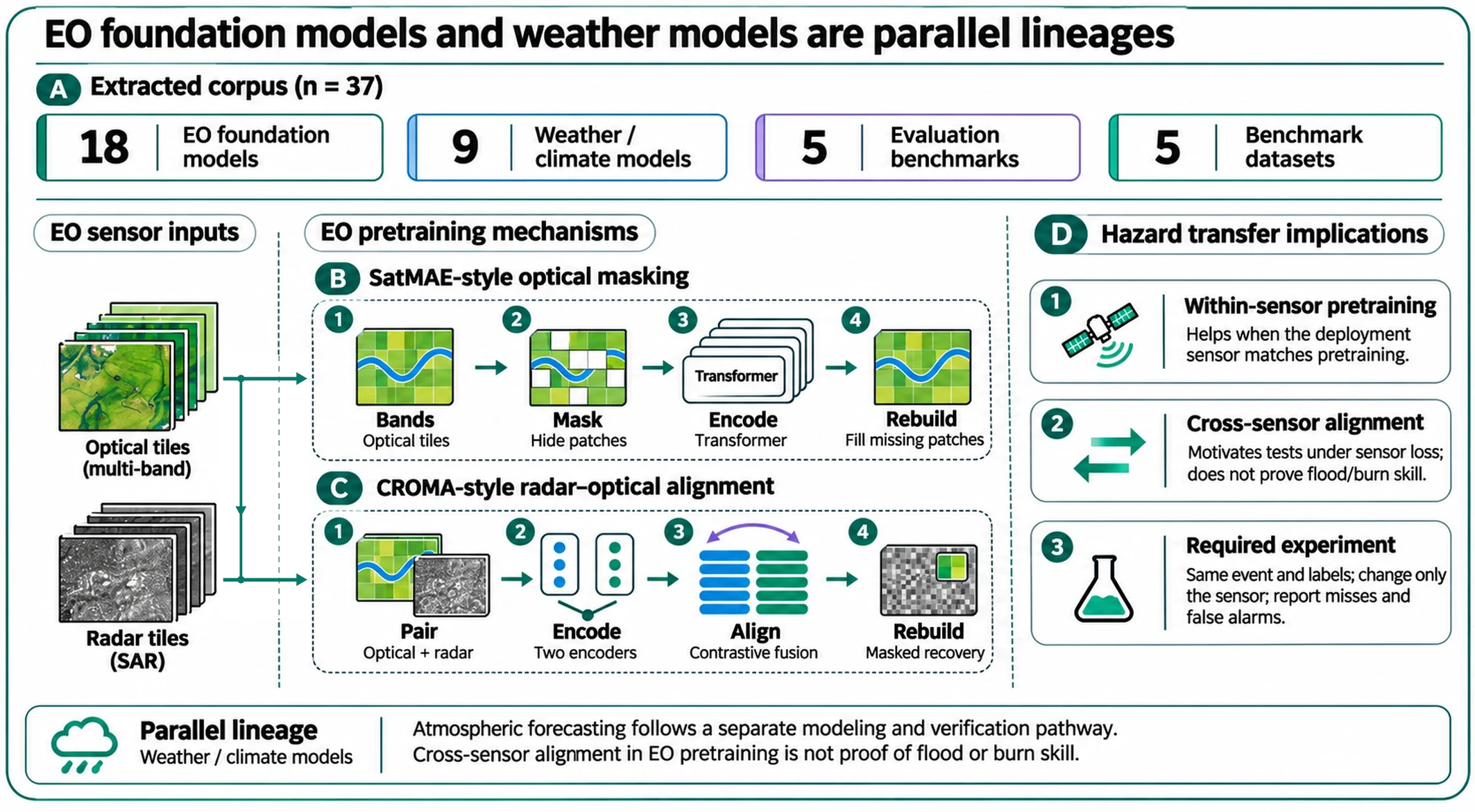}
\caption{EO foundation models and weather models are parallel lineages. (A) Extracted corpus composition. (B) SatMAE-style optical masking pathway. (C) CROMA-style radar--optical alignment pathway. (D) Implications for hazard transfer tests.}
\label{fig:fig2}
\end{figure*}

Figure~\ref{fig:fig2} summarizes two pretraining mechanisms, not a ranking of their hazard performance. SatMAE learns from masked temporal and multispectral satellite observations. Its representation is trained to recover missing information from the visible context. CROMA combines paired optical and radar encoders, cross-modal contrastive learning and masked reconstruction through multimodal fusion. These objectives encourage useful representations, but downstream flood or wildfire performance still depends on the adaptation data and evaluation setting \citep{satmae,croma}.

The distinction matters when the operational sensor configuration changes. Within-modality pretraining may support a downstream task whose observations resemble the training inputs. Cross-modal alignment creates a reason to investigate transfer when one modality is unavailable; it is not a measurement of that transfer. CROMA's separate encoders and fusion machinery are compressed into the schematic for readability. A test of one missing input, a test on a new satellite and a test of adaptation after acquiring target-sensor labels should therefore be reported separately.

Recent models broaden the range of inputs and adaptation strategies. DOFA conditions processing on spectral information to accommodate heterogeneous sensors; AnySat addresses varying modalities, resolutions and scales; TerraMind combines multimodal learning with generative capabilities; and Prithvi-EO-2.0 extends multitemporal Earth-observation pretraining \citep{dofa,anysat,terramind,prithvi2}. These are different design choices with different testable advantages. For a hazard user, the relevant question is whether a particular choice improves the specified task under matched labels, input availability and computational resources. Generated or inferred modalities must not be treated as independent measurements of the environment.

Table~\ref{tab:tab3} organizes representative models and benchmarks by what they contribute and what still needs to be tested. This replaces a cross-paper accuracy leaderboard with a comparison of mechanisms and evidence requirements. It also avoids treating model size, released embeddings and downloadable model weights as interchangeable indicators of capability or reproducibility.

\begin{table*}[!t]
\centering
\small
\caption{Representative model mechanisms and the evidence needed for hazard use}
\label{tab:tab3}
\begin{tabular}{p{0.22\textwidth}p{0.35\textwidth}p{0.35\textwidth}}
\toprule
\textbf{Model or benchmark} & \textbf{Reusable capability} & \textbf{Hazard-specific evidence still required} \\
\midrule
SatMAE (Cong et al., 2022) & Masked temporal and multispectral representation learning. & Event/region transfer and explicit input-availability tests. \\
CROMA (Fuller et al., 2023) & Separate radar and optical encoders, contrastive alignment and fused reconstruction. & Distinguish missing modality from unseen sensor; test each operational input set. \\
DOFA; AnySat (Xiong et al., 2024; Astruc et al., 2025) & Accommodation of heterogeneous spectral inputs, modalities or scales. & Matched label budgets and explicit tests on the intended sensors and resolutions. \\
Prithvi-EO-2.0; TerraMind (Szwarcman et al., 2026; Jakubik et al., 2025) & Multitemporal pretraining and multimodal or generative representations. & Ablations against strong task-specific models; provenance of any inferred modality. \\
GEO-Bench; PANGAEA (Lacoste et al., 2023; Marsocci et al., 2026) & Shared tasks and standardized model evaluation settings. & Verify that benchmark labels, split and metrics match the proposed hazard use. \\
Pangu-Weather; GraphCast (Bi et al., 2023; Lam et al., 2023) & Learned evolution of atmospheric fields. & Extreme-event verification at relevant leads, followed by downstream impact evaluation. \\
GenCast (Price et al., 2025) & Probabilistic atmospheric forecasts. & Calibration and decision value for the particular impact threshold and region. \\
NeuralGCM (Kochkov et al., 2024) & Hybrid learned components and atmospheric dynamics. & Verification at the temporal scale and process relevant to the application. \\
Aurora (Bodnar et al., 2025) & Adaptable Earth-system forecasting representations. & Task-specific extreme and transfer tests under operational information constraints. \\
\bottomrule
\end{tabular}
\tablenote{This is a selected methodological comparison, not a complete inventory of the 37-record contextual corpus. No numerical accuracy ranking across different tasks or protocols is implied.}
\end{table*}

\subsection{Atmospheric forecasting is a parallel pathway}
\label{sec:42}
Pangu-Weather and GraphCast illustrate advances in medium-range atmospheric prediction, while GenCast produces probabilistic forecasts, NeuralGCM combines learned components with atmospheric dynamics, and Aurora adapts a foundation-model approach to several Earth-system forecasting tasks \citep{pangu,graphcast,gencast,neuralgcm,aurora}. Their outputs and verification settings differ from those of an encoder adapted to segment flood water or burned terrain. Neither lineage is the final stage of the other.

Connecting the lineages could support impact-oriented warning: atmospheric forecasts inform hydrological, fire or exposure models, and Earth observations update the state of the land surface. However, each connection adds an uncertainty and timing requirement. A useful coupled test should trace forecast issue time, forcing uncertainty, spatial resolution, observation availability and the downstream decision threshold. The integration proposed by Reichstein and colleagues provides a research direction, not a reason to infer that a coupled system has already been validated for every hazard \citep{reichstein25}.

\subsection{A benchmark score is conditional evidence}
\label{sec:43}
GEO-Bench and PANGAEA organize diverse Earth-observation tasks and expose the importance of consistent adaptation settings and strong supervised baselines \citep{geobench,pangaea}. A benchmark result is conditional on its labels, split, decoder, tuning budget and metric. Differences among those choices can change model ordering. Corley and colleagues' audit of foundation-model reporting reinforces the need to examine the underlying comparison rather than borrowing a state-of-the-art label from another paper \citep{corley}.

For hazard applications, repeated use of a familiar event can also narrow the effective diversity of the evidence. Many papers evaluated on one held-out flood do not amount to tests on many independent floods. Dataset-level separation, event-level replication and pretraining provenance should be reported together. The scientific value of a new model is clearest when its claimed advantage survives a controlled comparison and remains visible across several relevant shifts, rather than depending on one favorable partition.

\section{Uncertainty, explanations and operational reliability}
\label{sec:5}
\subsection{Uncertainty reporting is not the same as calibration}
\label{sec:51}
\begin{figure*}[!t]
\centering
\reviewfig{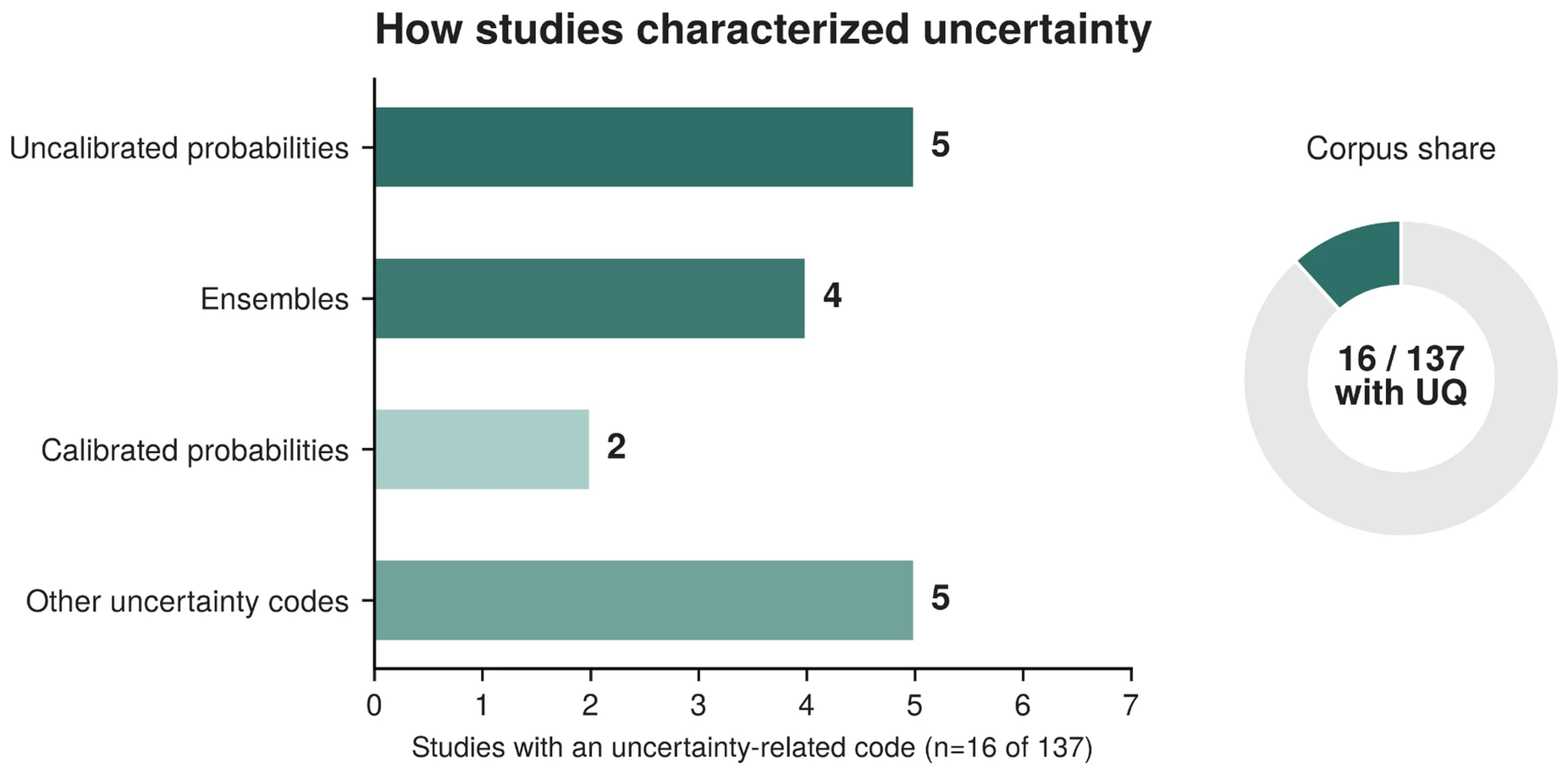}
\caption{Uncertainty-related reporting in the 137-study corpus.}
\label{fig:fig3}
\end{figure*}

Figure~\ref{fig:fig3} shows the uncertainty-related coding in the application corpus. Sixteen studies (11.7\%) had a recorded method: five used uncalibrated probabilistic outputs, four ensembles, two explicitly calibrated probabilities, and five other uncertainty-related approaches. The remaining 121 had no method recorded under this coding scheme. These categories describe reporting practices; they are not a score of the quality of every uncertainty estimate. In particular, an output between zero and one is not automatically a trustworthy probability.

Calibration asks whether events assigned a given probability occur at a corresponding frequency. Modern neural networks can require explicit calibration, and uncertainty behavior can deteriorate under distribution shift \citep{guo,ovadia}. For a flood map, reliability should therefore be checked on held-out events or regions at the thresholds used for action. For a warning, probability verification should be conditioned on lead time and event severity. An overall calibration statistic may conceal poor behavior for rare but consequential outcomes.

The required uncertainty summary depends on the product. Segmentation may need spatial error summaries and event-level confidence intervals; a peak-flow forecast may need prediction intervals, threshold exceedance probabilities and coverage by lead time. Ensemble spread can be informative, but agreement among models does not establish correctness when they share inputs, labels or structural biases. Calibration data must remain separate from final evaluation data. Resampling should also respect the independent unit, such as an event or basin, rather than treating nearby pixels as independent replicates.

\subsection{Explanations should answer a defined question}
\label{sec:52}
Explainability methods were recorded in 27 studies (19.7\%). A visually plausible saliency map or feature ranking can help diagnose a model, but it does not demonstrate that the explanation faithfully represents the learned relationship. Saliency sanity checks show why visual appeal alone is insufficient \citep{adebayo}. Explanations should be linked to a question: whether a model depends on an implausible feature, whether behavior changes across environments, or whether an input failure can be identified before use.

For impact studies, predictive explanation must also be separated from causal attribution. A relationship between vegetation and structure loss, for example, may be useful for risk screening without proving that changing vegetation alone would produce the predicted change in loss. Spatial validation and an explicit observational interpretation are important when translating explainable GeoAI into planning advice. The Palisades fire study provides a recent example of this distinction between model-based associations and intervention claims \citep{palisades}.

\subsection{Reproducibility and latency are part of the result}
\label{sec:53}
Explicit code availability was recorded in 52 studies (38.0\%). Table~\ref{tab:tab4} brings this result together with the other trust-related indicators. Reproducibility requires more than a model name: preprocessing versions, band definitions, normalization, label rules, split identifiers, adaptation settings and decision thresholds determine the experiment. FAIR principles provide a useful basis for organizing these resources so that other users can locate and reuse them \citep{fair}.

\begin{table*}[!t]
\centering
\small
\caption{Reporting and evaluation indicators in the 137-study application corpus}
\label{tab:tab4}
\begin{tabular}{p{0.20\textwidth}p{0.22\textwidth}p{0.28\textwidth}p{0.22\textwidth}}
\toprule
\textbf{Indicator} & \textbf{Studies} & \textbf{Share (\%)} & \textbf{Meaning of the indicator} \\
\midrule
Any out-of-distribution test & 98 & 71.5 & At least one event, region, time or sensor test reported somewhere in the study. \\
Any uncertainty-related method & 16 & 11.7 & Includes uncalibrated probabilities, ensembles and other variability/scenario methods. \\
Explicitly calibrated probabilities & 2 & 1.5 & A subset of the 16 uncertainty-related records. \\
Explainability method & 27 & 19.7 & A method was reported; faithfulness or causal interpretation is not implied. \\
Explicit code availability & 52 & 38.0 & A code availability statement or link was present in the coded paper. \\
Geospatial foundation-model fine-tuning & 10 & 7.3 & A named Earth-observation/geospatial foundation model was adapted to the task. \\
Headline random split & 48 & 35.0 & The discrete random-split headline code. \\
Headline sensor holdout & 2 & 1.5 & The discrete headline code; not a census of all sensor-related experiments. \\
\bottomrule
\end{tabular}
\tablenote{The denominator is 137 for every percentage. Indicators overlap; counts must not be summed. Publication-time availability coding is distinct from a present-day repository audit.}
\end{table*}

End-to-end latency is equally important. Inference time measures only one part of the path from an acquisition or forecast to a delivered product. Data access, quality control, atmospheric correction, registration, tiling, postprocessing and review can dominate the operational delay. A comparison intended for deployment should report the complete delay and the hardware and input conditions under which it was measured. Efficiency is a genuine advantage when it preserves the decision-relevant performance needed by a resource-constrained user; it is not a substitute for that performance.

\section{What the application audit establishes}
\label{sec:6}
\subsection{Model families and hazard coverage}
\label{sec:61}
\begin{figure*}[!t]
\centering
\reviewfig{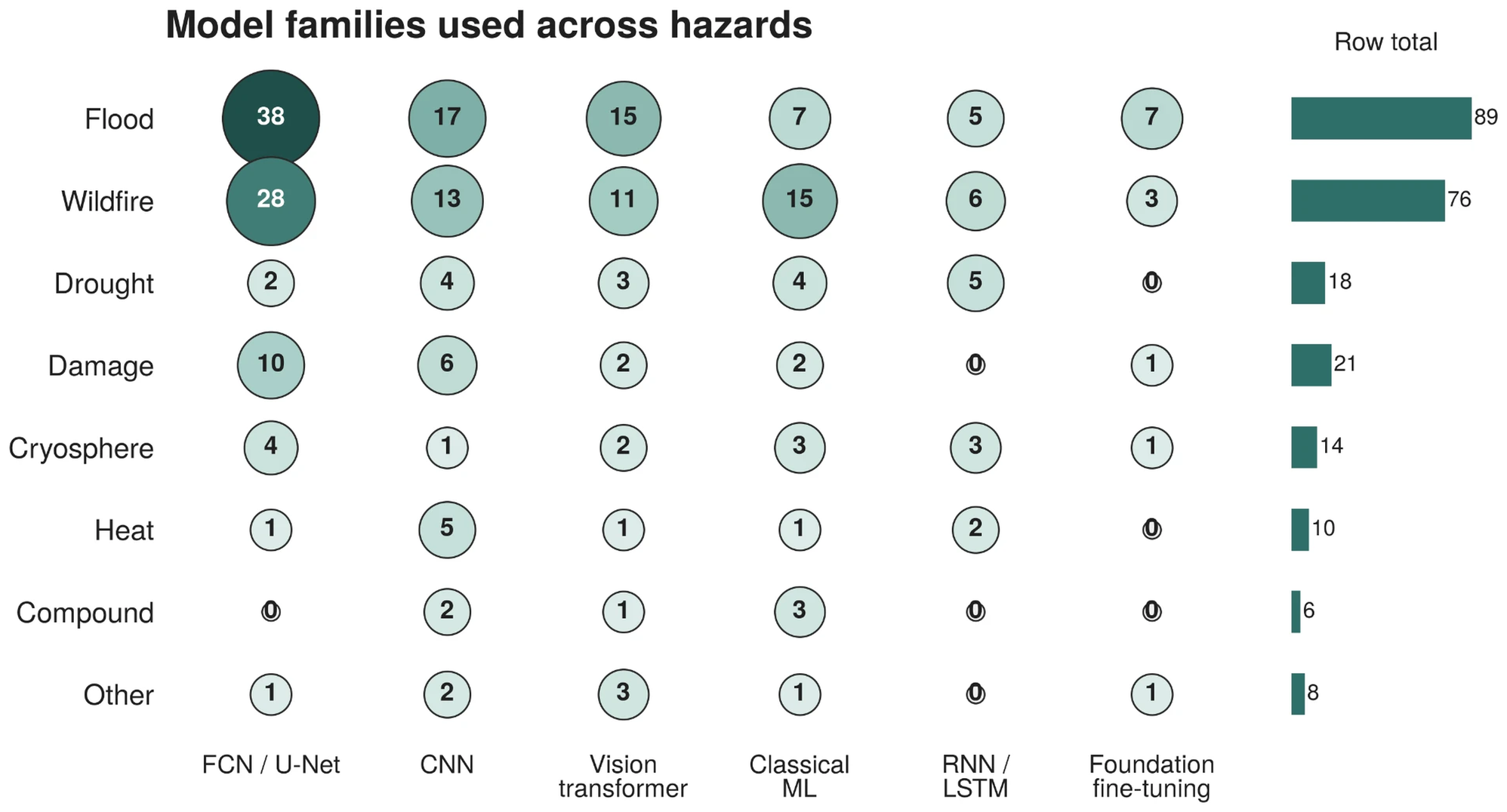}
\caption{Model families used across hazards.}
\label{fig:fig4}
\end{figure*}

The audit contains 56 flood, 51 wildfire, 16 drought, 14 damage, 11 cryosphere, nine heat, five compound-hazard and five other hazard labels. Because studies can receive several labels, these numbers describe coverage rather than exclusive categories. Figure~\ref{fig:fig4} shows the corresponding model-family cross-tabulation. Encoder-decoder segmentation was recorded in 71 distinct studies, convolutional networks in 38, transformers or vision transformers in 31, and classical machine-learning comparators in 29. Named geospatial foundation-model fine-tuning occurred in ten.

The larger row totals in the bubble matrix are counts of model-family tag assignments, not numbers of independent studies. A paper comparing U-Net, a transformer and a classical model contributes to several cells. The same distinction applies to hazards: a paper addressing both flood and damage assessment contributes to both rows. The figure describes the methodological mix of the corpus; it cannot establish that a common model family is more accurate or that a less frequent one is less suitable.

\subsection{Where the evidence is located}
\label{sec:62}
\begin{figure*}[!t]
\centering
\reviewfig{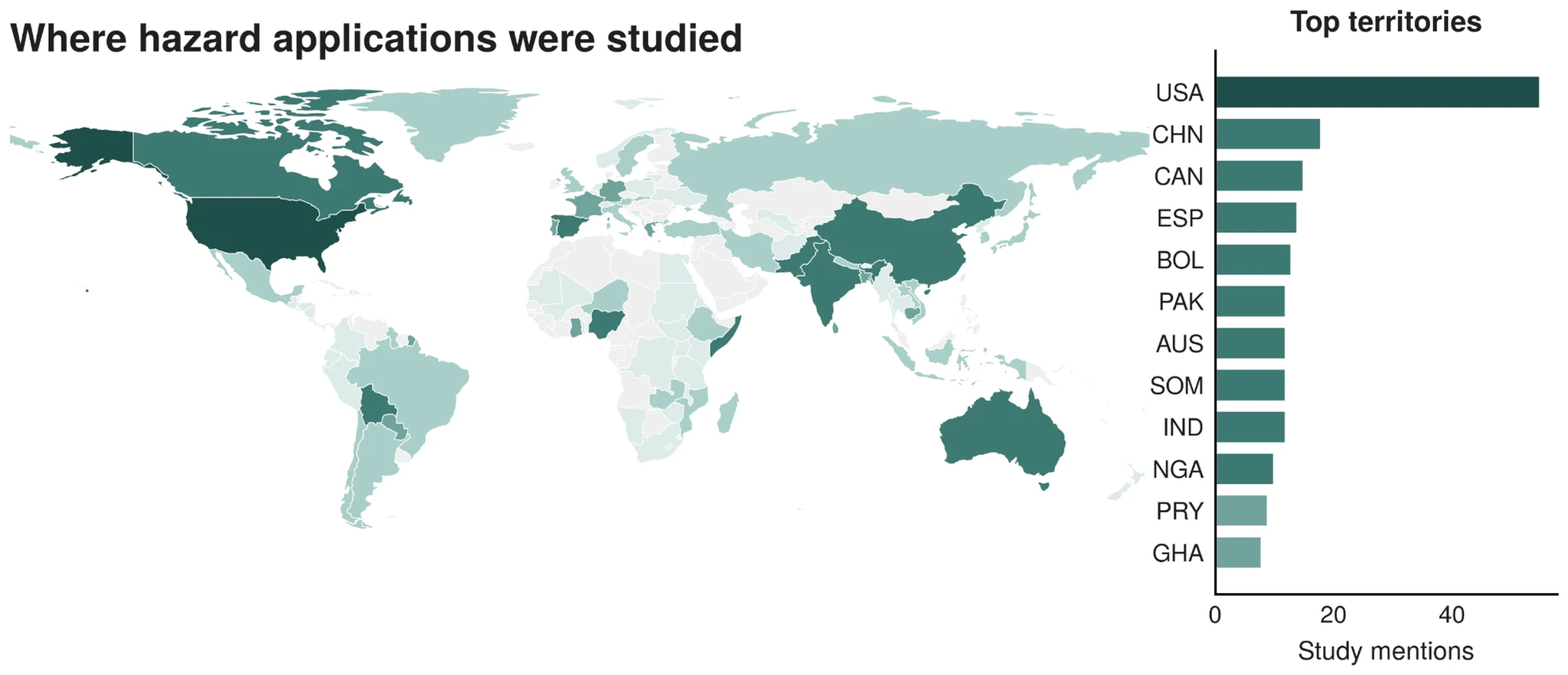}
\caption{Where hazard applications were studied.}
\label{fig:fig5}
\end{figure*}

Figure~\ref{fig:fig5} maps the locations recorded in the 110 geographically assignable studies. The United States accounts for the largest number of country mentions, with other territories represented unevenly. This distribution should be read alongside the prominence of flood and wildfire studies and the reuse of multinational benchmark datasets. A country can be represented through training data, test data or both; a filled polygon does not establish that the model has been independently evaluated across all environments in that country.

\begin{table*}[!t]
\centering
\footnotesize
\setlength{\tabcolsep}{3pt}
\renewcommand{\arraystretch}{0.88}
\caption{Hazard-task constraints and priority transfer tests}
\label{tab:tab5}
\begin{tabular}{p{0.19\textwidth}p{0.23\textwidth}p{0.28\textwidth}p{0.22\textwidth}}
\toprule
\textbf{Hazard (study labels)} & \textbf{Observations and target} & \textbf{Working limiting factor} & \textbf{Priority test} \\
\midrule
Flood (56) & Optical/SAR water mapping; rainfall and flow for warning. & Cloud, canopy and label definition for maps; forcing and lead time for forecasts. & New events/regions; sensor substitution; issue-time forecast verification. \\
Wildfire (51) & Thermal/optical active fire, burned area, severity; meteorology for danger/spread. & Observability for mapping; process evolution and weather for forecasting. & Separate targets; hold out fires and years; assess small-event misses. \\
Drought (16) & Soil moisture, vegetation condition and categorical drought references. & Reference definition, aggregation and forecast horizon. & Hold out years/regions; compare targets at a common spatial and temporal support. \\
Damage (14) & Very-high-resolution imagery, building or infrastructure references. & Label cost, scene heterogeneity and cross-disaster differences. & Hold out disasters/sites; state pre-event inputs and local adaptation. \\
Cryosphere (11) & Snow cover/SWE, glacier fronts and glacial lakes. & Sparse independent references and sensor/season ambiguity. & Regional and seasonal holdouts; reference uncertainty and acquisition conditions. \\
Heat (9) & Atmospheric fields, surface thermal observations and air-temperature references. & Forecast lead time and translation from fields to exposure. & High-impact event tests; lead-specific reliability and exposure-relevant targets. \\
Compound (5) & Paired event chains, such as fire followed by snow or hydrological change. & Coupled processes, time lags and aligned labels. & Withhold complete event chains; propagate uncertainty across stages. \\
\bottomrule
\end{tabular}
\tablenote{The constraints are testable hypotheses for specific tasks, not fixed properties of hazards. Hazard labels overlap. Five additional records carry other hazard labels. Representative evidence is discussed in Section 7.}
\end{table*}

The practical implication is to evaluate geographic applicability before extrapolating. Environmental similarity, sensor characteristics and label availability can differ substantially within the same country and across borders. An area-of-applicability analysis can help identify conditions unlike the training data, but it complements rather than replaces an independent test \citep{meyer21}. Reporting the actual evaluation locations and their environmental characteristics is therefore more informative than presenting a global map of data coverage as evidence of global readiness.

\subsection{Which shifts receive headline tests?}
\label{sec:63}
\begin{figure*}[!t]
\centering
\reviewfig{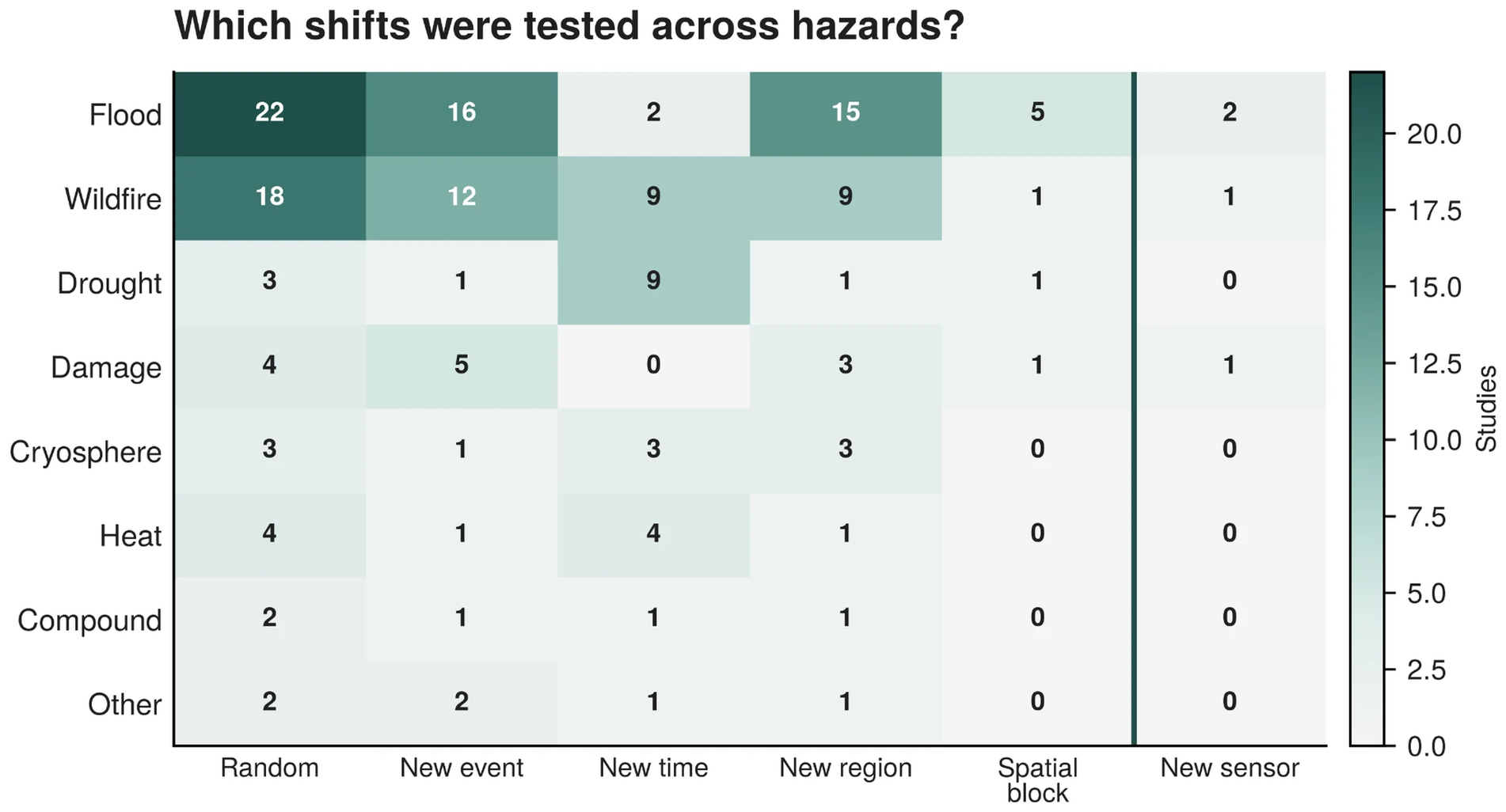}
\caption{Which shifts were tested across hazards?}
\label{fig:fig6}
\end{figure*}

Ninety-eight studies (71.5\%) reported at least one out-of-distribution experiment. The specific headline codes were random split in 48 studies (35.0\%), held-out event in 31 (22.6\%), held-out time in 27 (19.7\%), held-out region in 25 (18.2\%), spatial block in seven (5.1\%), and held-out sensor in two (1.5\%). Figure~\ref{fig:fig6} shows their distribution across hazards. These codes overlap, so they must not be read as a partition of the corpus or as successive levels attained by mutually exclusive groups.

Random splitting is the largest single displayed category, not a majority of the corpus. It can be appropriate for a question about interpolation within a well-defined data distribution. Its limitation arises when an interpolation result is used to support a claim about a new event, place or sensor. Conversely, merely reporting an extra transfer experiment does not establish adequate performance for a specified deployment. The audit records whether a test was present; the strength of its result still depends on the held-out conditions, reference quality and comparison design \citep{roberts}.

The rarity of headline sensor tests is the most concentrated gap in the displayed validation designs. It motivates targeted experiments under sensor change, especially where operational continuity depends on radar during cloud or on substituting one constellation for another. It does not show that all models fail under such changes, nor that only two papers in the wider literature have ever considered sensor differences. The inference concerns the principal test designs in this coded sample.

\subsection{Foundation-model studies and the comparison group}
\label{sec:64}
\begin{figure*}[!t]
\centering
\reviewfig{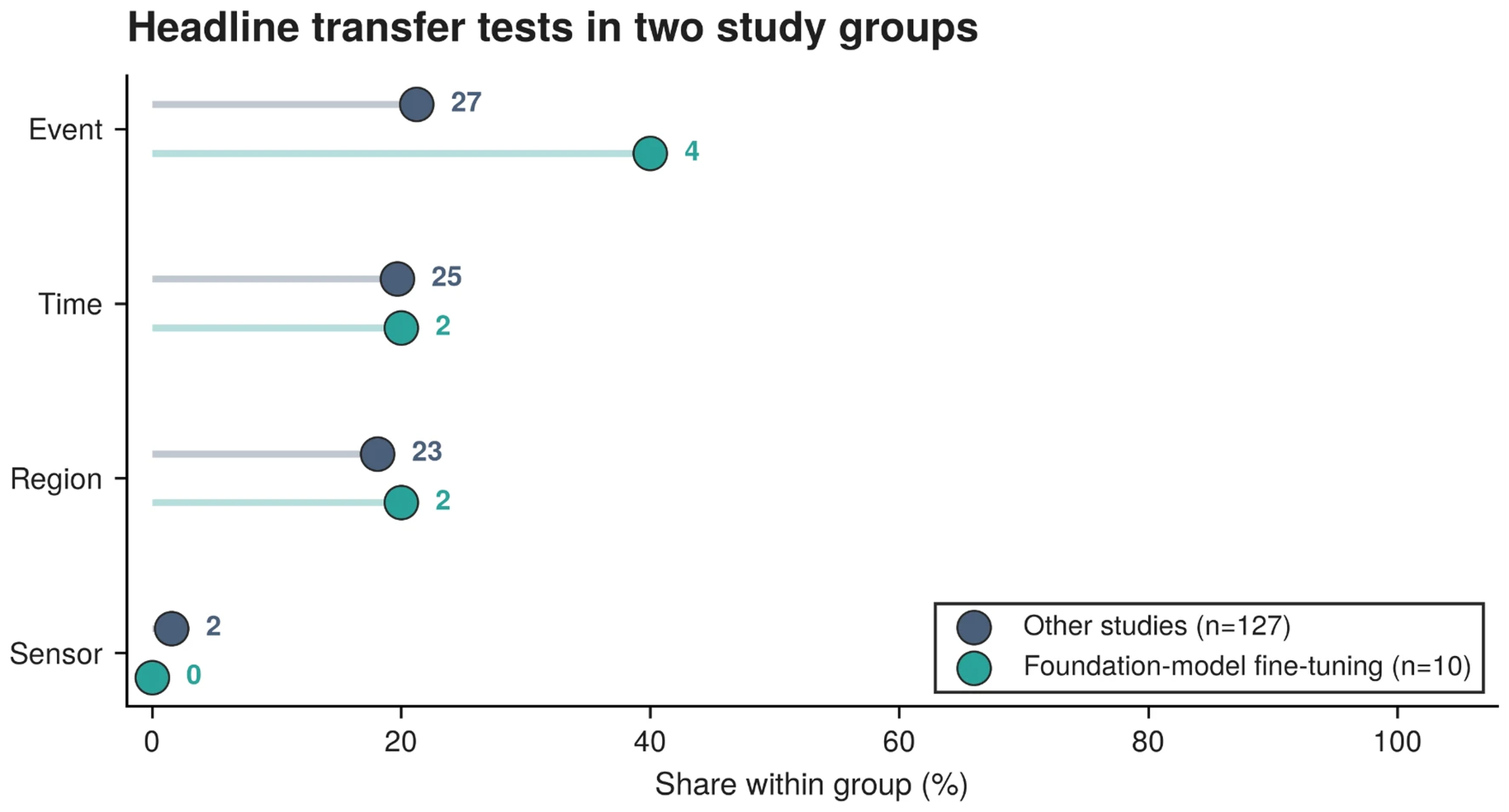}
\caption{Headline transfer tests in two study groups.}
\label{fig:fig7}
\end{figure*}

Figure~\ref{fig:fig7} uses within-group percentages rather than raw counts to compare the ten foundation-model fine-tuning studies with the other 127 applications. Event holdouts occurred in four foundation-model studies and 27 other studies; time holdouts in two and 25; region holdouts in two and 23; and sensor holdouts in zero and two. The corresponding event shares are 40.0\% and 21.3\%, but the groups differ greatly in size and composition. Their differences are descriptive, not evidence that fine-tuning causes a particular validation practice.

The zero sensor-holdout count in the fine-tuning group is an evidence gap, not an estimate of the probability that a foundation model will transfer successfully. Establishing comparative transfer performance requires matched experiments, ideally spanning several events and sensor conditions. The audit therefore points to a research priority rather than a verdict on the model class.

\section{Different hazards require different tests}
\label{sec:7}
Table~\ref{tab:tab5} links the hazard coverage to observations, targets and limiting factors. These factors are working hypotheses about a hazard-task pair, not permanent properties of an entire hazard. The same wildfire can generate an image-segmentation problem, a spread forecast and a post-fire hydrological assessment, each with different information requirements.

\subsection{Flood: mapping water and anticipating discharge}
\label{sec:71}
Flood applications provide the clearest contrast between representation learning and forecasting. Sen1Floods11 has enabled shared flood-mapping experiments, while Kuro Siwo expands multi-temporal SAR observations and manually annotated events \citep{sen1,kuro}. These datasets support reproducible comparisons, but the scope of a result remains tied to the labels and event selection. Cloud-aware optical systems further show that operational mapping involves acquisition availability and end-to-end processing, not segmentation alone \citep{portales}.

Prithvi-based studies illustrate why the baseline must receive comparable attention. Early work assessed a geospatial foundation model on flood inundation mapping, while U-Prithvi combined a pretrained encoder with a U-Net architecture \citep{li,uprithvi}. In the latter study's strengthened-baseline comparison, Bolivia mean IoU was 87.77\% for U-Net and 87.70\% for U-Prithvi. This near equality does not invalidate pretraining; it shows that the conclusion changes when the task-specific comparator is strengthened. A model's advantage should therefore be attributed to the tested configuration, not to the foundation-model label in isolation.

A recent compact-model preprint similarly compared pretraining, random initialization, frozen features, U-Net and computational cost under a common water-mapping task. Its results support an accuracy-efficiency trade-off rather than universal superiority of a pretrained encoder \citep{compact}. For warning, the relevant evidence is different: forecasting studies such as global prediction in ungauged watersheds evaluate future streamflow rather than a satellite-observed water mask \citep{nearing}. A mapping benchmark cannot substitute for basin and event holdouts at useful warning lead times.

\subsection{Wildfire: distinguish detection, aftermath and spread}
\label{sec:72}
Active-fire detection identifies an ongoing thermal or spectral signal; burned-area mapping identifies a changed surface; severity and structure-loss analyses characterize consequences. Fire-spread and danger forecasting concern future conditions. Conflating these targets can make a method appear to solve a wider operational problem than it actually addresses. Models for danger forecasting that combine dynamic environmental information with geographic context illustrate why weather and location must be treated jointly rather than replaced by a larger optical encoder (Shams Eddin et al., 2023).

Event-separated datasets are valuable, but the label-generation process remains part of the task. The Sentinel-2 active-fire benchmark covering 25 California wildfires assigns each wildfire to a single partition and provides explicit metadata and rule-derived masks \citep{mitra}. Such a split prevents the same wildfire from appearing in training and testing; it does not turn a spectral rule into an independent observation of every fire boundary. Evaluation against separately reviewed or independently observed targets is needed when the claim extends beyond reproducing the labeling rule.

For structure-loss screening, pre-event inputs, spatial separation and the distinction between association and causation are especially important \citep{palisades}. A decision about treatment or land-use change needs more evidence than a model explanation, even when predictive discrimination is useful. Operational wildfire comparisons should report misses in difficult settings, such as small fires or complex interfaces, as well as aggregate segmentation or classification scores.

The Kincade Fire vineyard analysis provides a related landscape-scale example: multisource observations can characterize wildfire impacts and exposure at field scale, but a spatial association between land cover and burn severity is not by itself a transferable estimate of a management intervention's effect \citep{compact}. This distinction matters when remote-sensing evidence is used to support adaptation planning across agricultural and urban landscapes.

\subsection{Drought: the reference defines the forecast}
\label{sec:73}
Drought is expressed through several related but distinct targets, including soil moisture, vegetation condition and categorical drought assessments. DroughtCast forecasts the United States Drought Monitor, whereas vegetation-health and soil-moisture forecasting studies predict different environmental quantities \citep{droughtcast,lees,lesinger}. A model can be skillful relative to one reference without reproducing another. Reported accuracy therefore depends partly on the target definition, spatial support, aggregation and forecast horizon.

Temporal holdouts are necessary for an operational forecasting claim, but they should be supplemented by tests of rare severity classes and geographic transfer when those uses are intended. Inputs should be screened for availability at issue time, including the release timing of reanalysis and monitoring products. A useful experiment compares the effect of changing the reference target with the effect of changing the model, while retaining a common split and evaluation scale. This tests the label-definition hypothesis without assuming that target choice always dominates architecture.

\subsection{Heat: from atmospheric skill to exposure}
\label{sec:74}
Heat applications link atmospheric forecasts, near-surface air temperature, land-surface temperature and human exposure. These variables are related but not interchangeable. Satellite thermal observations provide surface information; ground observations and atmospheric models are needed to interpret the conditions experienced by people. Next-day canopy urban-heat-island prediction that combines satellite and ground measurements illustrates the value of defining the exposure-relevant target explicitly \citep{heat}.

Tests on recent high-impact heat events show that strong average weather-forecast performance does not ensure equally strong performance in an extreme case \citep{pasche}. Such results identify a limitation of the evaluated systems and lead times; they do not, on their own, establish a fundamental physical predictability limit. Operational heat evaluation should stratify errors by intensity, lead time and relevant population or urban setting, while retaining an appropriate forecasting baseline. The nine heat labels in the audit identify a small evidence subset, not a measure of the societal importance of heat risk.

\subsection{Cryosphere: sparse references and changing observation conditions}
\label{sec:75}
Snow-water-equivalent estimation, glacier-front delineation and glacial-lake mapping have different reference and observation requirements. Regional transfer of a long short-term memory model for snow-water equivalent addresses a different question from adapting an image encoder to delineate lakes or glacier fronts \citep{ssl4sar,glacial}. Independent regions, acquisition conditions and reference sources should therefore be identified explicitly, rather than summarized by a generic cryosphere accuracy claim.

SSL4SAR reports a 293 m mean calving-front distance error on the CaFFe dataset. On a separate multi-annotator subset, its ensemble result was 75 m, compared with 38 m disagreement among human annotations \citep{ssl4sar}. The subset comparison indicates remaining model error relative to expert agreement; the different evaluation populations must not be merged into one model-versus-human comparison. Annotation variability is a relevant reference property, not proof of an unavoidable error floor.

\subsection{Damage and cascading impacts}
\label{sec:76}
Damage mapping often confronts changes in imaging conditions, building form and disaster context. Work on transferring a global building-mapping model to heterogeneous post-disaster imagery demonstrates that the extraction problem itself can shift before any downstream damage classification is attempted \citep{hu}. A held-out disaster should therefore preserve the operational target and make clear whether local labels, pre-event images or sensor-specific adaptation were used.

Cascading hazards add a temporal connection between processes. Burn severity can influence subsequent debris-flow hazard, and fire-related canopy change can alter snow persistence. Studies connecting pre-fire conditions, burn severity and debris-flow exposure, and recent multisource work on wildfire severity and snow persistence, illustrate these linked pathways \citep{wells,creek}. A credible cascade test must follow the sequence of events and propagate uncertainty across stages. High accuracy for the first map does not establish the skill of the downstream impact forecast. The five explicitly compound-hazard studies in the audit motivate shared event-chain benchmarks rather than conclusions about the frequency of cascading risk itself.

\section{Choosing the test from the decision backward}
\label{sec:8}
\subsection{Specify the claim before selecting the split}
\label{sec:81}
The proposed protocol begins with an action, not an architecture. State who will use the output, what decision it informs, when that decision is made and which information is available by then. Specify the target, spatial support, relevant event severity, acceptable delay, and the costs or operational consequences of misses and false alarms. The evaluation should then reproduce the change that could invalidate that decision: a new event, region, season, sensor or input-availability condition.

Table~\ref{tab:tab6} links these claims to concrete tests and reporting requirements. The 'transfer ladder' in Figure~\ref{fig:fig8} is an organizing device, not a universal ranking of difficulty. A well-designed temporal forecast test can be more demanding than a geographically distant but environmentally similar image test. Likewise, a sensor test does not replace an event test. The required combination follows from the deployment claim.

\begin{strip}
\footnotesize
\setlength{\tabcolsep}{3pt}
\renewcommand{\arraystretch}{0.88}
\refstepcounter{table}%
{\let\sffamily\rmfamily
\noindent{\bfseries\sffamily\tablename~\thetable}\par
\noindent{\sffamily Decision-backward tests linked to the headline audit counts}\par
}%
\vspace{0.25\baselineskip}%
\label{tab:tab6}
{\centering
\begin{tabular}{p{0.20\textwidth}p{0.22\textwidth}p{0.28\textwidth}p{0.22\textwidth}}
\toprule
\textbf{Deployment question} & \textbf{Headline studies (\%)} & \textbf{Evaluation requirement} & \textbf{Decision-facing reporting} \\
\midrule
Interpolation within a sampled distribution & 48 (35.0) & Random split with a clear sampling unit and no broader transfer claim. & Error by relevant class and environmental condition. \\
Use in spatially separated areas & 7 (5.1) & Withhold spatial blocks; state block scale and separation. & Spatial variation, failure locations and applicability. \\
Use in a future season or period & 27 (19.7) & Withhold time; restrict inputs to those available at issue time. & Lead-time error, reliability and temporal drift. \\
Use during a new hazard event & 31 (22.6) & Withhold complete events from fitting, tuning and calibration. & Event-level misses, false alarms and uncertainty. \\
Use in a new region & 25 (18.2) & Withhold a region and document environmental and label differences. & Regional performance and conditions outside applicability. \\
Use after sensor or modality change & 2 (1.5) & Specify unseen sensor, ablation or adaptation; match labels and budgets. & Sensor-conditional error, availability and full data-to-product latency. \\
\bottomrule
\end{tabular}\par
}%
\tablenote{Counts use the discrete headline codes in Figure~\ref{fig:fig8}; random cross-validation and spatial cross-validation are not silently added. Codes overlap. For all transfer claims, retain a matched task-specific/operational baseline, document pretraining overlap and assess uncertainty in the deployment setting.}
\end{strip}

\subsection{Experiment 1: maintain mapping under sensor change}
\label{sec:82}
Consider an agency that must continue flood mapping when cloud prevents an optical acquisition. A controlled experiment would define the same target event and reference labels, match the adaptation-label and tuning budgets, and compare the available sensor configurations. A task-specific baseline should be retained alongside the pretrained model. Observation time, spatial resolution and the information carried by each sensor should be aligned where possible and otherwise reported as part of the operational comparison.

Two questions should be separated. A controlled sensor comparison asks how performance changes when the observing configuration changes under closely matched conditions. An operational substitution test asks which available product provides the best decision support when perfect input parity is impossible. Neither should be described as zero-shot sensor transfer if target-sensor labels or model updates are used. Pretraining overlap and any target-domain adaptation must be disclosed.

Thresholds should be selected before final testing. Report missed inundation, false alarms, event-level performance and uncertainty, and stratify failures by conditions relevant to response, such as cloud, forest, dense development and narrow water features. Report the delay from data availability to the delivered map. A gain in average overlap is useful only when it does not obscure a failure mode that prevents the intended inspection or response decision.

\subsection{Experiment 2: warn before the peak}
\label{sec:83}
For a warning service, freeze the information set at each issue time and evaluate storms or water years that are excluded from model fitting, tuning and calibration. Historical data should be reconstructed as they would have been available operationally, rather than supplemented with later observations or retrospectively revised products. The baseline should reflect the service's actual alternative, such as an operational forecast, a hydrological model or an appropriate simpler statistical method.

Evaluate peak magnitude, arrival time, threshold exceedance, usable lead time and probabilistic reliability. Report misses and false alarms by event severity, not only average error across many uneventful periods. The comparison must preserve the same issue times and admissible inputs for competing methods. For a coupled weather-to-impact system, propagate atmospheric uncertainty through the downstream component instead of evaluating a single best atmospheric trajectory as though it were certain.

\subsection{Evaluate whether pretraining changes the decision}
\label{sec:84}
A minimal attribution experiment compares the pretrained architecture with a randomly initialized counterpart, a strong task-specific model, and the operational or classical baseline relevant to the action. Match input bands, spatial and temporal support, labels and tuning budgets; document differences that cannot be matched. Vary the available downstream labels to test label efficiency rather than inferring it from one training set. Repeated seeds quantify training variability, whereas repeated held-out events quantify a different source of variability; both should be identified rather than pooled indiscriminately.

The practical output is a trade-off among decision-relevant errors, lead time, cost and coverage. A compact model may be preferable when it supports deployment on limited hardware, while a larger model may be justified where a reproducible gain changes the action. The result should establish the condition under which the model helps, not simply declare a winner across settings that were never jointly evaluated.

\section{Discussion and research priorities}
\label{sec:9}
\subsection{From model capability to evidence of useful transfer}
\label{sec:91}
The audit identifies an imbalance between the breadth of potential deployment conditions and the specificity of the tests reported as headline evaluations. This is not an argument against foundation models. It is an argument for experiments that can identify their contribution. A reusable encoder is valuable when it improves a defined task, reduces the labels or resources required, or extends reliable coverage under a meaningful shift. Those outcomes should be demonstrated directly.

Four research priorities follow. First, build event-diverse and sensor-explicit benchmarks that distinguish observed water, inundation, active fire, burned area and impact targets. Second, report pretraining provenance alongside downstream holdouts so that representation reuse can be separated from evaluation independence. Third, evaluate calibration and failure modes under the same shifts as accuracy. Fourth, test coupled weather and Earth-observation systems at the issue times and lead times required by an actual warning decision. These priorities complement broader calls for integrated early warning and transparent benchmarking \citep{reichstein25,corley}.

\subsection{Test the limiting-factor hypotheses}
\label{sec:92}
The framework proposes that representation learning is most useful when the observations contain the needed signal and the principal difficulty is learning a transferable representation from limited downstream labels. This predicts an advantage that should be measured across label budgets and independently held-out environments. When visibility is the dominant limitation, adding an informative observation may matter more than enlarging an encoder. When future process evolution is the challenge, forcing information and forecast lead time may dominate a mapping architecture. When a decision depends on calibrated probabilities, a higher overlap score alone is insufficient.

These propositions require intervention-style comparisons, not categorical declarations that one constraint always controls a hazard. Change the sensor while holding the event and reference as stable as possible; change the lead time while preserving the forecast task; or change the reference index while preserving the model and partition. The resulting performance differences can reveal which limitation is most important in that setting. Several constraints can bind simultaneously, and an improvement in one may expose another.

\subsection{Transfer lessons from agricultural and urban monitoring}
\label{sec:93}
Adjacent environmental applications help clarify these evaluation choices without being counted as additional hazard studies. The Sentinel-2 crop-yield review highlights the need to interpret imagery alongside crop development and the intended prediction scale \citep{yield}. Drone-based and satellite time-series studies of branched broomrape illustrate a different target-identification problem: an observed crop-stress signal must be connected to the reference condition and observation timing \citep{drone,broom}. These studies concern biotic stress, not direct validation of drought or climate-extreme prediction. Their relevance is methodological: visually similar stress responses should not be assigned the same cause without supporting evidence.

Spatial mapping studies provide another useful distinction. Farmland extent and visible-boundary mapping from NAIP imagery concern observable land-cover geometry, not a legal cadastral boundary \citep{farmland}. Optical urban-canopy mapping and resilience screening connect a remotely sensed feature to a planning context, but do not turn canopy detection into a validated heat-exposure forecast \citep{canopy}. These examples reinforce the need to define the output before selecting the model, preserve scene or geographic independence when that is the deployment question, and separate a mapped proxy from the decision variable it is intended to inform.

\subsection{Scope of the evidence and implications for use}
\label{sec:94}
The descriptive percentages characterize the accessible, coded corpus. Full-text availability, the search source and the distribution of hazards influence its composition. The figures do not estimate the worldwide prevalence of a practice or provide causal comparisons among model families. The small fine-tuning and sensor-holdout subsets also make broad group-level conclusions inappropriate. A transparent interpretation is to use the audit to identify specific missing tests and to use the application literature to design those tests.

The same principle applies to climate relevance. A flood or fire map can support adaptation without demonstrating climate-change attribution. A short-range weather forecast is not a climate projection. An association with present-day damage is not evidence of a mitigation benefit. Keeping these distinctions explicit makes the review useful to researchers and practitioners alike: it identifies what an output supports, what additional evidence an application requires, and where a promising method deserves a controlled operational comparison.

\begin{strip}
\centering
\reviewfig{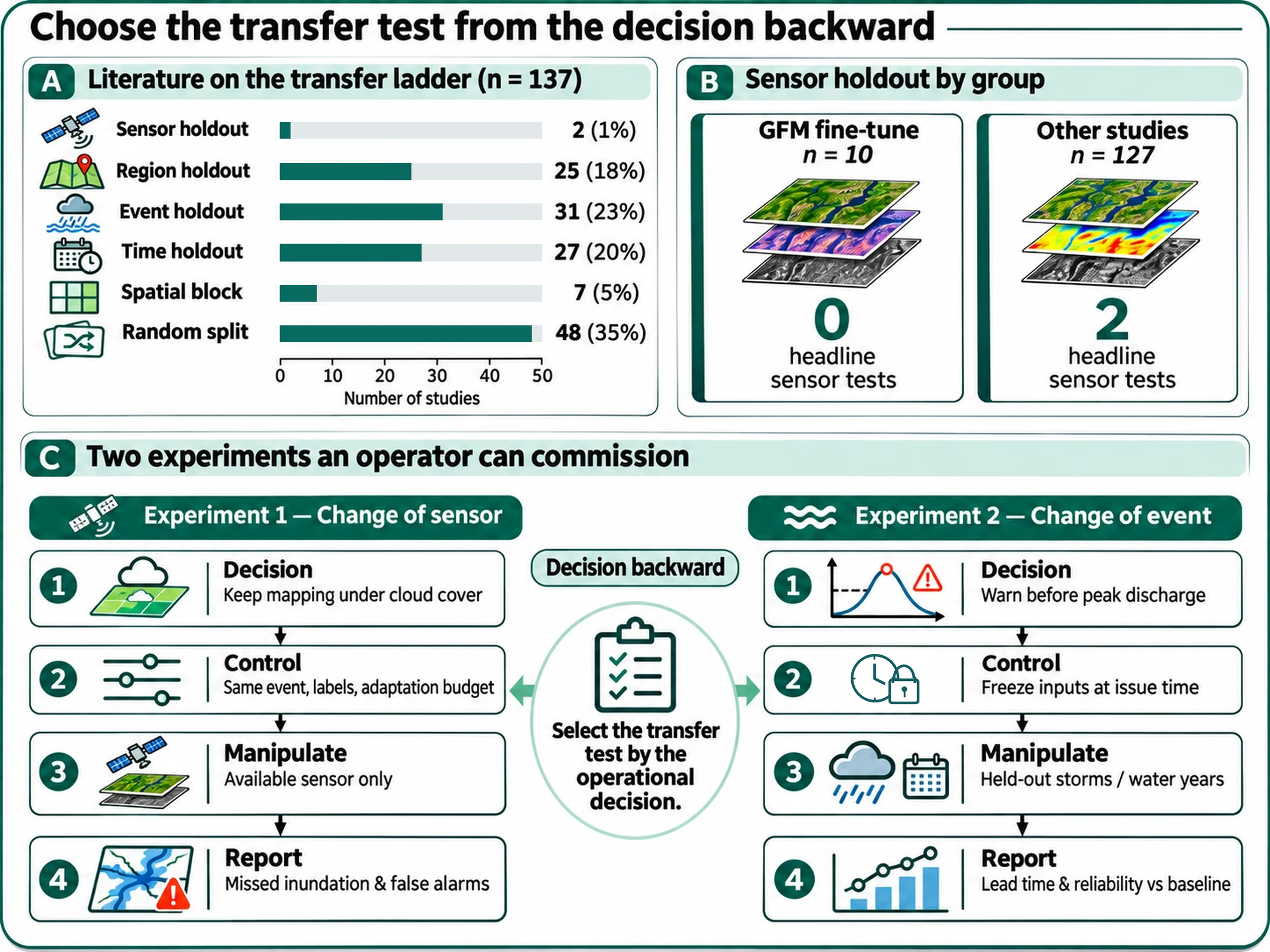}
\captionof{figure}{Choose the transfer test from the decision backward.}
\label{fig:fig8}
\end{strip}

\section{Conclusions}
\label{sec:10}
Geospatial foundation models expand the ways Earth-observation information can be reused, but climate resilience requires evidence about transfer under the conditions in which decisions are made. The 137-study audit shows meaningful use of out-of-distribution experiments alongside uneven coverage of specific headline shifts. Held-out events and regions were recorded in 31 and 25 studies, respectively; sensor holdout appeared in two, with none among the ten foundation-model fine-tuning studies. Uncertainty-related reporting was also limited, and only two studies were explicitly coded for calibrated probabilities.

These findings define an evaluation agenda, not a verdict that pretrained models are ineffective. Earth-observation representation learning and atmospheric forecasting offer complementary capabilities, and their integration should be tested through information availability, uncertainty propagation and decision timing. Across hazards, label definitions, observability, forecast horizon and operational requirements determine what improvement is possible and how it should be measured.

The central recommendation is to choose the test from the decision backward. Define the action, reproduce the relevant shift, compare models under matched information and adaptation budgets, and report the errors, reliability and latency that govern practical use. Progress is demonstrated when a model improves a decision under those conditions, rather than when a familiar benchmark score is presented as evidence of readiness for an unfamiliar world.

\section*{Data availability statement}
\label{data-availability-statement}
Screening and coding materials supporting this review are available from the corresponding author upon reasonable request. Licensed full-text publications are not redistributed.

\section*{Author contributions}
\label{author-contributions}
{\raggedright
PF: Conceptualization, Methodology, Investigation, Data curation, Formal~analysis, Software, Visualization, Writing -- original draft, Writing -- review \& editing.\\
MN: Conceptualization, Methodology, Investigation, Data curation, Formal~analysis, Software, Visualization, Project administration, Writing -- original draft, Writing -- review \& editing.\par
}

\section*{Funding}
\label{funding}
This research received no specific grant from any funding agency in the public, commercial, or not-for-profit sectors.

\section*{Conflict of interest}
\label{conflict-of-interest}
The authors report no competing interests to declare.

\section*{Ethics statement}
\label{ethics-statement}
The study uses remote-sensing, environmental, infrastructure, parcel and aggregated public statistical data. It involves no human participants and no identifiable individual-level records.

\section*{Generative AI statement}
\label{generative-ai-statement}
During the preparation of this work, the authors used ChatGPT to improve grammatical accuracy, refine sentence structure and enhance visualizations. All AI-generated revisions were thoroughly reviewed and edited by the authors to ensure relevance and accuracy. The authors take full responsibility for the scientific content, interpretation, citations and final presentation.

\renewcommand{\bibsection}{%
  \par\vspace{0.6em}%
  {\sectionfont\large\bfseries References}\par\nopagebreak\vspace{0.35em}%
}
\balance
\bibliographystyle{IEEEtranN}
\bibliography{references}
\end{document}